\documentclass{sigcomm-alternate}

\usepackage{cite}
\usepackage{algorithmicx}
\usepackage{algorithm2e}
\usepackage{stfloats}
\usepackage{fixltx2e}
\usepackage{graphicx}
\usepackage{epstopdf}
\usepackage{subfigure}
\usepackage{color}
\usepackage{url}

\def\eg{{\it e.g., }}
\def\ie{{\it i.e., }}

 \date{}
\begin{document}

\title{Opportunities in a Federated Cloud Marketplace}

\numberofauthors{3}
\author{
\alignauthor Hamed Haddadi\\
        \affaddr{Queen Mary University of London, UK}\\ 
\alignauthor Georgios Smaragdakis\\
        \affaddr{T-Labs/TU Berlin, Germany}\\
\alignauthor K. K. Ramakrishnan\\
        \affaddr{WinLab, Rutgers University, NJ, USA}\\
}


\maketitle

\begin{abstract}
Recent measurement studies show that there are massively distributed hosting
and computing infrastructures deployed in the Internet. Such infrastructures
include large data centers and organizations' computing clusters. When idle,
these resources can readily serve local users.  Such users can be smartphone or
tablet users wishing to access services such as remote desktop or CPU/bandwidth
intensive activities. Particularly, when they are likely to have
high latency to access, or may have no access at all to,
centralized cloud providers. Today, however, there is no global marketplace
where sellers and buyers of available resources can trade. The recently
introduced marketplaces of Amazon and other cloud infrastructures are limited
by the network footprint of their own infrastructures and availability of such
services in the target country and region.
In this article we discuss the potentials for a federated cloud marketplace 
where sellers and buyers of a number of resources, including storage, 
computing, and network bandwidth, can freely trade. This ecosystem can be 
regulated through brokers who act as service level monitors and auctioneers. 
We conclude by discussing the challenges and opportunities in this space.

\end{abstract}




\section{Introduction}\label{sec:introduction}

Internet applications are becoming more complex and more resource-intensive. To
cope with the increasing requirements and complexity of these applications,
applications providers are continuously adding computational, storage, and
network resources in the Internet. Some of the application providers are now
among the largest infrastructure providers in the Internet~\cite{Cartography}.
Application requirements also dictate how application or infrastructure
providers deploy and expand their footprint in the Internet. Some of them rely
on large data centers located in strategic network locations, colocated with
large Internet exchange points (IXPs) or in large metropolitan areas \eg
Amazon, Rackspace, Equinix, ThePlanet. Others rely on a number of well
distributed data centers \eg Google, Yahoo!, Microsoft, Limelight. Some others
deploy highly distributed servers deep inside the network, \eg Akamai.

For example, Akamai operates more than $100,000$ servers in more than $2,000$
locations across nearly $1,100$
networks~\cite{ImprovingPerformanceInternet2009,Akamai-Network}. Google is
reported to operate tens of data centers and front-end server clusters
worldwide~\cite{MovingBeyondE2E2009,Tariq:What-if}.  Microsoft has deployed its
CDN infrastructure in 24 locations around the world~\cite{WindowsAzure}.
Amazon maintains at least 5 large data centers and caches in at least 21
locations around the world~\cite{amazon}. Limelight operates thousands of
servers in more than 22 delivery centers and connects directly to more than 900
networks worldwide~\cite{LLNetworks}. Equinix has presence in 24 data centers
in North and South America, 10 data centers in Europe and 5 data centers in
Asia-Pacific region \cite{equinix}. 

Data centers are often deployed to satisfy local markets. A rough estimation of
the currently operated small to medium size datacenters accounts more than 2200
co-located data centers in around 84 countries \cite{datacentermap}.  
The increasing popularity of data-intensive open-source
applications, such as Hadoop \cite{Hadoop}, to perform data analytics is
driving the deployment of private data centers.  Prime operators of such data
centers are financial companies, large corporations, web service providers,
just to name a few. The exact number of such data centers is unknown, but
we expect that it is quite significant.  
Furthermore, a large number of scientific
computational clusters and data centers hosted in universities and research
centers also contributes to the penetration of such datacenter infrastructures.
These
infrastructures are typically operated by a single authority and are either
dedicated to the applications run by the infrastructure owner or are leased to
third parties. As the profit margins of such infrastructures go down, we 
anticipate
that the trend of leasing storage and computational resources to third parties
will inevitably increase.

The deployment and operation of storage, computation, and network
infrastructures comes at a cost.  Among the highest contributing factors are
the cost of energy \cite{CostOfCloud,AkamaiCutting:2009}, especially cooling,
network bandwidth \cite{Akamai-Network}, and the administration cost that may
even exceed the cost of the hardware \cite{CoupledPlacement}. To amortize these
costs, some of the infrastructure providers have leveraged the
advances that resource virtualization offers to lease resources to tenants, \ie
third parties applications. In some cases, this provides a substantial revenue
flow, see for example the Amazon Web Services. Others follow this paradigm as
their core business, for example Rackspace or ThePlanet. 

However, the offers for resources is currently limited to the footprint of each
infrastructure provider. Such a footprint may not be agile enough for the
placement and operation of all the applications. This set up may not be optimal
for mobile users, those with high latency or low bandwidth to major cloud
providers, or those in under privileged countries. Moreover, potential buyers
of resources can not freely express the requirements for resources for their
applications as the configuration and presence of infrastructures may be
limited.  Recent studies \cite{EmbarassinglyDistributed} confirm our
observations that a more agile allocation of resources may benefit a number of
important applications.  Another limitation is that there is no common
interface to lease cloud and network resources from multiple cloud providers.
This leads to custom solutions that cannot scale to different providers, and
thus, increase the overhead to lease resources as well as increases the burden
of new cloud providers to enter the market.In this paper we present a first
attempt at theoretical analysis of the feasibility of a \emph{federated}
marketplace for cloud computing resources were resources are available for
leasing under a common brokerage scheme, and we sketch the protocol design
principles to materialize it. 

Most of the work on incentives for leasing or sharing Internet resources has
focused on peer-to-peer
systems~\cite{Cohen2003,P2P-backup,p2p-MarketDesign-EC10,ECHOS}. Recently,
researchers have investigated pricing schemes for virtual resources, but the
focus is on releasing resources in a single
datacenter~\cite{PriceIsRight,aboveclouds}. Work on incentives has also
considered unused bandwidth resources
\cite{InvisibleHand-key-paper,InvisibleHand} and Shapley value has been used as
a tool to allocate the right value for an Internet resource
\cite{CooperativeSettlement:ToN}. The community project
Seattle\footnote{\url{https://seattle.cs.washington.edu/html/}} encourages
individuals to altruistically donate a fraction of their computing power to
others by way of peer-to-peer networking. In~\cite{Wood:2012}, Wood et. al.
present CloudNet, a a cloud platform architecture that utilizes virtual private
networks to link cloud and enterprise sites in a dynamic and adaptive manner,
although this approach is still dependant on cooperation between cloud service
providers and does not necessarily focus on independent, \textit{casual}
clouds.


\section{Marketplace: Incentives, Requirements, and Case Studies}\label{sec:scenarios}

We argue for creation of a global resource marketplace where sellers and buyers
of resources can trade on the specifications and price of the leased resources.
First we highlight the rational behind building a federated marketplace. We then
describe the requirements to enable such a cloud marketplace. We conclude with
the presentation of some promising case studies.

\subsection{Incentives}\label{sec:marketplace-incentives}

A global marketplace for resources increases the awareness of the available
resources to a wider group of potential buyers. A potential buyer of resources
can have access to resources where it needs them, when it needs them. The entry
cost for the introduction of smaller sellers and buyers is also lower. This
enables healthy competition and contributes to a wider spectrum of prices and
offered services. It is also possible to gain access to resources very close to
end-users or target groups that may not be available in the current resource
marketplace.

Today, many data center resources are under-utilized. This is due to the fact
that the operators of resource infrastructures provision for the peak. By taking
advantage of daily cycles, it is possible to offer un-utilized resources at
attractive prices. Infrastructure providers derive additional revenue and
application providers have access to reduced costs for their deployment.

The buyer of resources have more degrees of freedom when expressing their needs
as they are not limited to the offers of a limited number of large providers. It
also enables a better negotiation of the quality and the price of resources.
This can be a positive as the price of different resources may significantly
differ among providers and locations \cite{CloudCmp}, due to their specific
footprint.

The advances of virtualization technology allows the fast reservation of
resources and also allows applications to expand or shrink on-demand, over
timescales of minutes. This elasticity can significantly reduce operational
costs of applications and enable more flexible charging models, such as the
typical pay-as-you-go policies enabled by cloud services~\cite{aboveclouds}.

The enhanced performance of commercial applications leads to higher revenue
\cite{amazon-study-about-transaction-speed}.  We foresee a unique opportunity
for infrastructure and application providers to jointly increase their revenue
as well as the performance of applications. It can also be a catalyst for the
introduction of new applications, hence lowering the barrier for innovation.

\subsection{Requirements}\label{sec:marketplace-requirements}

To enable such a resource marketplace, it is imperative to overcome a number of
challenges. In the following, we outline a number of necessary conditions that
must be satisfied to ensure these challenges are addressed. In later section of
this paper we propose a number of incentives to ensure that these conditions are
satisfied.

{\bf Truthfulness of sellers:} The sellers have to unveil the true location of
their resources, the type and utilization of each resource, and a suggested
price per unit of usage in each location and time-of-day for each resource. The
broker's rating system and measurements of Quality of Experience (QoE) will
enforce such truthfulness. 

{\bf Truthfulness of buyers:} The buyers of resources should truthfully unveil
the real demand for a resource and the real value for the different resources
in different locations and time-of-day. 

{\bf Robust billing:} The broker has to ensure that the best match between
supply the demand of a resource takes place. The allocation of resources has to
be a revenue maximization mechanism for the involved parties while leaving
resources available for future requests.  Negotiations between the involved
parties has also to be allowed until a consensus on the price is met.  Other
soft incentives such as economies of scale should also be present~\cite{cloudbill}

{\bf Mature and privacy-preserving technology:} Resource virtualization is the
prominent technology to support the lease of a number of resources. Some
privacy concerns may arise when running virtualization in the wild, in
untrusted environments and with multiple tenants. We do not address the privacy
issues in this paper~\cite{5189563}.
 

\subsection{Case Studies}\label{sec:marketplace-case-studies}

To exemplify the utility of a cloud marketplace, we present a number of case-studies:

{\bf Cloud-assisted Smartphones.} Increasingly, people use tablets and
smartphones to perform daily tasks or access their main desktop. For most
individuals, with the uptake of netbooks, many applications will not run on the
device natively anymore. Access to these applications using remote desktop
services requires high bandwidth and more importantly low latency. Our proposed
market place can be useful for such scenarios. 

{\bf Online Gaming.} Some metropolitan areas host a number of data centers,
either commercial or scientific ones.  The peak hour for these data centers is
typically during the day, with periods of low utilization in the evening, when the
end-users of the data centers are going home.  This is the time when the home
entertainment peak hour begins. Many entertainment applications are interactive
and their performance depends significantly on the end-to-end delay. A prominent
example is this of online games. The operators of online games would likely
prefer to use data centers very close to the end users to minimize latency and
avoid a build-out of special-purpose data centers (\eg using GPUs) to overcome
the effect of large latencies to their primary data center. Medium or even micro
data centers that are under-utilized can be used to host such applications and
significantly improve the end-user experience. Our marketplace can enable this.

{\bf Sport events.} Global-scale events such as the Olympic games can take
advantage of the massively distributed infrastructure to scale the delivery of
real-time or recorded video. They can also take advantage of the time-of-day
effect in order to lease low cost resources and significantly reduce the
delivery cost by leasing resources is non-peak local hours (lead the sun, not
follow it).  Our marketplace is a catalyst for this.

{\bf The midnight stock exchange analysis.} There are a number of data centers
in universities and research centers in metro areas such as London, Frankfurt
etc.  Many studies show that the cost of deploying commercial data centers can
be significantly higher in these metropolitan locations. On the other hand, data
analytics applications are key for stock forecasting. The need to analyze
financial data is expected to grow. Through our proposed marketplace, the above
mentioned data centers can be leased at low prices when they are not used;
mainly at night or when they are idle.

{\bf CDNs can also lease their resources as well.} CDNs utilize thousands of
servers to cope with the demand for content. Still, during the non-peak hour,
the utilization of such infrastructures is low. CDNs can advertise the available
resources using our marketplace and have an additional flow of revenue. 

\section{Participants and their roles}\label{sec:players}

\begin{figure}[t!]
\begin{center}
\includegraphics[width=1.1\columnwidth]{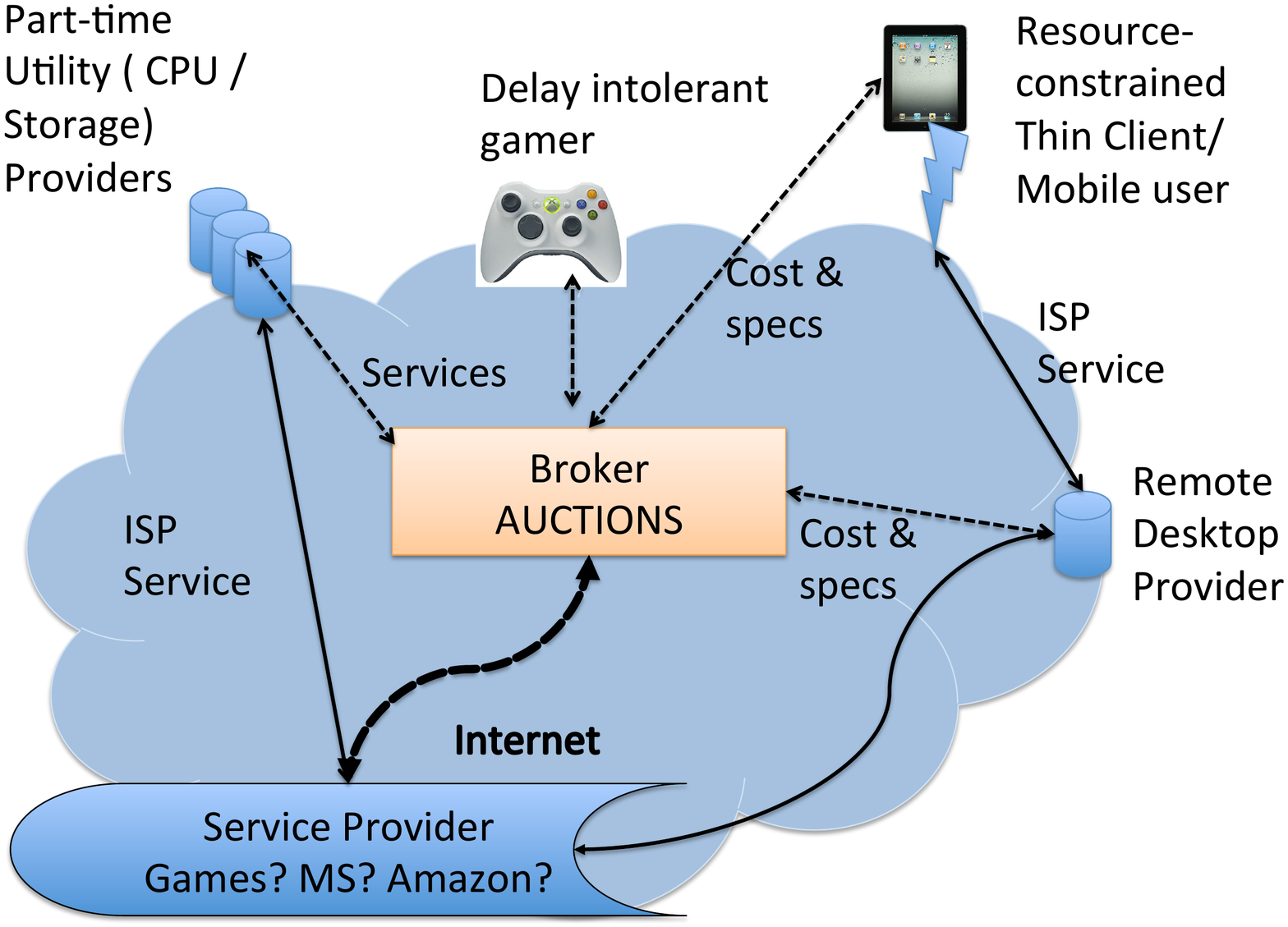}
\end{center}
\caption{The players in the cloud market ecosystem.}
\label{fig:system}
\end{figure}

Figure~\ref{fig:system} shows the participants and their interaction in our
cloud marketplace ecosystem. In this section we will expand on the individual
requirements of these roles.

\paragraph{Customer} The customer can be an individual, a small community of
users (\eg a local games store in developing countries), or a small business.
This customer may have access to a thin client or a tablet device, needing to
access a remote desktop service (\eg a version of Microsoft office). Industry predictions are that more than half the web
traffic will soon be on tablet devices~\cite{ciscopredict}. The incentive of
using a local cloud resource for the user would be that the user needs to invest
less on resources, while still having low-latency and convenient access on
resource constrained device like tablet.  Moreover, it will enable lower delay
when compared to accessing large cloud services located further away, and also
likely to be at a lower cost.

\paragraph{Network Providers} The Network or the Internet Service Provider (ISP)
delivers connectivity at a cost. The rise in Traffic Engineering (TE) and Smart/Sponsored Data Pricing~\cite{sen2013survey} can increase the effect of the ISP
on the market. However in this paper we assume that the ISP simply provides adequate
connectivity to local cloud resources, and we don't assume any further role for them.

\paragraph{Resource Providers (RP)} An RP is a cloud operator, or an
organization which can provide CPU, storage, memory or remote desktop licensed
software (\eg MS Office instances). Resource Providers can act as a relay for
thin clients and provide a range of services and applications (apps). The RPs
will define their base rates for auctions and Quality of Service (QoS)
guarantees. Their aim is to maximize their profit and utilization of their
resources. Using the proposed federated cloud application marketplace, the RPs
can make use of their temporary capacity and idle resources by leasing their
space resources for financial gains. 

\paragraph{Market/App Broker}  A broker acts as the provider of the mobile/cloud
app store,  auctioneer, and reliability score keeper, without having a
centralized control over the market. The broker receives requests and matches
them to providers, runs auctions between RPs and completes the bidding actions
on specified intervals. The broker maintains reliability scores and builds a
reputation system for different RPs and active directory and resource discovery
for their type of services, while setting criteria and assigning coefficients to
different attributes of importance, \eg memory and CPU. The broker collects
feedback and provides reputation metrics and resolves conflicts.

\section{Discussion}\label{sec:conclusion}

In this article we argued for a federated cloud-based resource marketplace. The
proposed architecture in this paper is aligned with demands and services needed
by users that typically have to depend on large-scale cloud offerings such as
those provided by the Amazon Elastic Cloud and Web services. This trend is
motivated by the increasing processing and bandwidth demands of applications
such as online games as well, where companies reach for the cloud\footnote{\url{http://online.wsj.com/article/SB10001424052702303343404577517074213415592.html}}
to enhance the user's visual experience and minimize delays. Our suggested
approach can meet the needs of environments where latency to access large cloud
services may be so large as to restrict the kinds of applications that can be
used.

This is an early attempt to examine the feasibility of an open market for
cloud-based resources. Our proposed resource marketplace aims to enhance the
exploration of the resources that are currently available but unknown to the
potential buyers, improves the agility of leasing resources in the Internet,
lowers the burden for introducing new players in the resource ecosystem, and
creates opportunities for new revenue flows and better services. In future work
we will investigate the proposed hypothesis by obtaining real usage data from cloud services in
order to take factors such as time of day and day of week into consideration
when leasing resources.  We also aim to improve the auctioning mechanism by use
of different strategy for different resources, \eg by matching customers with
the next highest resource cloud service providers if their top resource
requirement cannot be fulfilled.

\vspace{0.1in}
{\par\noindent\bf Acknowledgements}. 
\small{This work was done while Hamed Haddadi was at AT\&T Labs Research, on an EPSRC IT-as-a-Utility industrial secondment (EPSRC Ref EP/K003569/1). We acknowledge constructive feedback from Felix Cuadrado on earlier versions of this work.}

 \bibliographystyle{abbrv}
\bibliography{paper}

\end{document}